\documentclass[10pt,twocolumn]{article}
\usepackage[textwidth=7in,textheight=8.5in]{geometry}
\usepackage{graphicx}
\usepackage{amsmath} 
\usepackage{amssymb} 
\usepackage{fancyvrb} 
\usepackage{soul}
\usepackage{fancyhdr}
\usepackage{hyperref}
\makeatletter
\renewcommand{\maketitle}{\bgroup
\begin{flushleft}
  \begin{Huge}
  \textbf{\@title}\\
  \end{Huge}
  \vspace{1cm}
  \@author
\end{flushleft}\egroup
}
\makeatother

\title{Pion Fluctuation Study in Pb-Pb Collision at $2.76 TeV$ per nucleon pair from ALICE Experiment with Chaos and Complex Network-based Methods}
\author{%
    \textbf{{\large Susmita Bhaduri}}$^{1}$, \textbf{{\Large Anirban Bhaduri}}$^{2}$, \textbf{{\Large Dipak Ghosh}}$^{3}$\\
    $^{1,2,3}$Deepa Ghosh Research Foundation, Kolkata-700031,India \\
    \underline{$^{1}$susmita.sbhaduri@dgfoundation.in}\\
    \underline{$^{2}$bhaduri.anirban@dgfoundation.in}\\
    \underline{$^{3}$dipak.ghosh@dgfoundation.in}
}
\begin{document}
\twocolumn[
  \begin{@twocolumnfalse}
    \maketitle
  \end{@twocolumnfalse}
  ]
\noindent

\date{\today}

\begin{abstract}
Chaos and complex-network based study is performed to look for signature of phase transition in Pb-Pb collision data sample at $2.76 TeV$ per nucleon pair from ALICE Collaboration. The analysis is done on the pseudorapidity($\eta$) values extracted from the data of ALICE experiment and the methods used are \textit{Multifractal-Detrended-Fluctuation-Analysis(MF-DFA)}, and a rigorous chaos-based, complex-network based method - \textit{Visibility-Graph(VG)} analysis. The fractal behavior of pionisation process is studied by utilizing MF-DFA method for extracting the Hurst exponent and Multifractal-spectrum-width to analyze the scale-freeness and fractality inherent in the fluctuation pattern of $\eta$. Then VG method is used to analyze the fluctuation from a completely different perspective of complex network. This algorithm's scale-freeness detection mechanism to extract the \textit{Power-of-Scale-freeness-of-Visibility-Graph(PSVG)}, re-establishes the scale-freeness and fractality. 
Earlier, it has been shown that the scaling behavior is different from one hadron-nucleus($\pi^{-}$-AgBr(350 GeV)) to one nucleus-nucleus($^{32}$S-AgBr(200 A GeV)) interaction which is of comparatively higher total energy~\cite{Bhaduri20167}. In this work, we have compared the fluctuation pattern in terms of $3$ rigorous parameters - Multifractal-spectrum-width, Hurst exponent and PSVG, between Pb-Pb($2.76 TeV$ per nucleon pair) data and either of $\pi^{-}$-AgBr(350 GeV) or $^{32}$S-AgBr(200 A GeV) data, where both the interaction data are of significantly less energy than the ALICE data. We found that the values of the $3$ parameters are substantially different for ALICE data compared to the other two interaction data. As remarkably different value of long-range-correlation indicates phase-transition, similar change in the fluctuation pattern in terms of these parameters can be attributed to a phase-transition and also the onset of QGP.
\end{abstract}

\textbf{PCAS Nos.:} \textit{24.60.Ky, 24.60.Lz, 24.85.+p.}
\section{Introduction}
\label{intro}
The quantum chromodynamics (QCD)-the theory of strong interactions, within the Standard Model of the high-energy physics predicts that the nuclear matter goes through a process of phase transition to a state of de-confinement of quarks and gluons, at a critical temperature of about $160 MeV$~\cite{Borsanyi2010}, associated to an energy density of about $0.7 GeV/fm^3$~\cite{Karsch2001}. Shuryak had suggested long back that, by studying matter in this phase referred as Quark Gluon Plasma (QGP) phase~\cite{Shuryak1978} where quarks and gluons are not in confined state forming hadrons, one can have deeper insight into the basic properties of QCD matter in its normal state of confinement and also in the breaking process of chiral symmetry~\cite{Lee1974}.

After the introductory experimental efforts at LBNL and JINR in the 1970s, the theoretical and experimental research progressed with increasing energy at different laboratories like GSI SIS, BNL AGS and CERN SPS. Then in 2000, an analysis of the SPS heavy-ion experiment was done~\cite{Heinz2000,cern2000} which, by manifesting some of the important properties predicted for the QGP, showed the production of a new state of matter in Pb-Pb central collisions at $\sqrt{s_{NN}} = 17.3 GeV$. Then there was further progress with the heavy-ion programme at the BNL RHIC~\cite{Arsene2005,Adcox2005} and the CERN SPS~\cite{Specht2010}, which confirmed and fine-tuned the initial SPS results. The study of QCD at high temperatures embarked on a new age of precision measurements in 2010 with the start-up of LHC furnishing Pb-Pb collisions at the phenomenally high collision energies having magnitude far more than previously achieved.

The experimental analysis and decoding of the QGP state should provide more insight into deterministic and explainable properties of QCD, that are key to interpret hadron and nuclear characteristics. Different experimental observables have been optimized for probing dynamic evolution of the system and characterization of the various stages of the collision, to draw out the features of the produced matter. 
Conventionally, the initial state of a heavy-ion collision has been described utilizing the so-called Color Glass Condensate-CGC framework~\cite{Kovner1995,Kovchegov1997,Krasnitz1999,Krasnitz2000,Lappi2003,Lappi2006}, where an over-occupied ensemble of soft gluons has been presented up to the so-called saturation scale and above that the states are nearly unoccupied. The progression of the system from that state towards hydrodynamic and ultimately fully thermalized state could be delineated by very weak and strong couplings respectively, using kinetic theory and numerical holography~\cite{Chesler2011,Kurkela2015}. 
Although various attempts have been made, it has not been possible to detect the QGP directly. Hence it has been tried to obtain information about the initial state and the collision history, mostly from the final state hadrons measured in various experiments.

At the LHC experiments, there has been the huge increase of energy, up to $\sqrt{s_{NN}} = 5.02 TeV$ for Pb-Pb collisions. The outcomes are expected to provide more advantageous conditions of energy density and temperature leading to the production of a denser, hotter, longer-lasting medium. So, one of the goals of the high-strength experiments at LHC, namely ALICE, ATLAS, CMS, and LHCb, has been measuring the parameters more precisely utilizing the high strength of LHC and thus characterizing the new state of matter. The increase of the collision energy at LHC gave rise to huge increase of the production of hard probes, giving access to a new set of observables and it has been the most crucial impact of LHC~\cite{Foka2016}.

Another advantage at LHC has been the highly broad phase-space coverage attained by the detector systems in pseudorapidity-$\eta$ space. A phase transition to QGP state in a multiparticle production phenomena, normally gives rise to a remarkable change in the fluctuation process of an observable parameter like $\eta$ than that in the confined state.
Heat capacity of a system also changes noticeably when it undergoes phase transition. However, the energy density remains a steady function of the temperature. So, change in the pattern of multiplicity fluctuation in high-energy interactions, may be suggestive of QGP generation.

Bialas and Peschanski~\cite{bialash986} introduced in a new concept called intermittency, to study large fluctuations. It has been observed that multipion production in heavy-ion interaction shows a power-law behavior of the factorial moments with respect to the size of phase-space intervals in decreasing mode. An indication of a self-similar fluctuation is thereby obtained, which in turn indicated the fractal behaviors in statistical and geometrical systems. 
The study of fractal behavior of multipion production from the perspective of intermittent fluctuations using the method of factorial moment, had been an initial area of interest. There is a simple relationship between the anomalous fractal dimension and intermittency indices\cite{bia1ash1988,dewolf1996}. The cascading mechanism inherent in the multipion production process produces a fractal structure as a natural consequence. 

In the recent past, numerous techniques based on the fractal theory have been implemented to analyse the process of multipion emission~\cite{hwa90,paladin1987,Grass1984,hal1986,taka1994}. The most popular of them have been Gq moment and Tq moment developed by Hwa and Takagi~\cite{hwa90,taka1994}. Both these methods have been extensively applied to analyse the multipion emission process, after considering their merits and demerits~\cite{dghosh1995a,dghosh2012}. Then techniques like the Detrended Fluctuation Analysis(DFA) method~\cite{cpeng1994} have been introduced for detecting monofractal scaling parameters and the Hurst exponent, which is associated with fractal dimension inferred from the DFA function of a time-series~\cite{kantel2001}. This method has been used extensively for detecting long-range correlations in noisy and non-stationary time-series data~\cite{MSTaqqu1995,zchen2002}.  
Kantelhardt et al.~\cite{kantel2002} extended DFA to analyse non-stationary and multifractal time-series as a generalized version of DFA known as the multifractal-DFA (MF-DFA) method. 
The DFA and MF-DFA parameters have been used extensively for analyzing nonlinear, non-stationary data series for identifying their long-range correlations.
The multifractal nature of the distribution of shower particles around central rapidity region of Au-Au collisions at $\sqrt{s}_{NN}=200$A GeV, using MF-DFA method, was analysed by Zhang et al.~\cite{YXZhang2007}.
A number of multifractal analysis of particle production processes has been also reported in the recent times~\cite{Albajar1992,Suleymanov2003,Ferreiro2012,pmali2015}.

Recently new approaches have been proposed to study complex systems in terms of complex networks, as they provide us a quantitative model for large-scale natural systems(in the fields of biology, physics and the social sciences). The topological properties of these complex networks derived from the real systems provide useful information about the characteristics of the system.
Albert and Barab{\'{a}} have studied the latest advances in the field of complex network and examined the analytic tools and models for random graphs, small-world and scale-free networks, in the recent past~\cite{Albert2002,Barabasi2011}. Havlin et al. have discussed the application of network sciences to the analysis, perception, design and repair of multi-level complex systems which are detected in man-made and human social systems, in organic and inorganic matter, and in natural and anthropogenic structures~\cite{Havlin2012}. 

Lacasa et al. have introduced Visibility Graph analysis~\cite{laca2008,laca2009} method which has gained importance due its entirely different, rigorous approach to assess fractality. Lacasa et al. have analysed real time series in different scientific fields, using fractional Brownian motion(fBm) and fractional Gaussian noises(fGn) series as a theoretical framework. The Hurst parameter calculated for fractional Brownian motion(fBm) with different methods, often yields ambiguous results, because of the presence of non-stationarity and long-range dependence in fractional Brownian motion(fBm). Lacasa et al. mapped fractional Brownian motion(fBm) and fractional Gaussian noises(fGn) series into a scale-free Visibility Graph having the degree distribution as a function of the Hurst exponent~\cite{laca2009}. This way, they have applied classical method of complex network analysis to quantify long-range dependence and fractality of a time series~\cite{laca2009}. 
The accuracy of Visibility Graph algorithm has been established for artificial data and real data series by Lacasa et al. and their details are given in \cite{Hausdorff1996,Goldberger2002}. 
This method has been used productively for analyzing various biological signals in recent works \cite{Bhaduri2014,Bhaduri20163,nil2016,bhaduriJneuro2016,Bhaduri2017}. 
In view of the above, recently, using Visibility Graph method, we have analysed multiplicity fluctuation in ($\pi^{-}$-AgBr(350 GeV)) and ($^{32}$S-AgBr(200 A GeV)) interactions~\cite{Bhaduri20167,Bhaduri20171,Bhaduri20183}, the fractality of void probability distribution in ($^{32}$S-AgBr(200 A GeV)) interaction~\cite{Bhaduri20163,Bhaduri20165,Bhaduri20166,Bhaduri20172}, analyzed the azimuthal anisotropy in ($^{32}$S-AgBr(200 A GeV)) interaction~\cite{Bhaduri20182} and also proposed a new perspective of clan model of particle production process using this complex network scenario~\cite{Bhaduri20181}.

Zebende et. al. have studied long-range correlations resulting from temperature-driven liquid-vapor-phase transition in distilled water, using DFA method and have shown that the scaling exponent increases with temperature proceeding towards transition temperature~\cite{Zebende2004}. Zhao et. al. have studied the fractal properties of magnetization time series using MF-DFA and Visibility Graph method and confirmed Hurst exponent as a good indicator of phase transition for a complex system~\cite{Zhao2016}. 
It's evident from the experiments for liquid-vapor-phase transition~\cite{Zebende2004} and for magnetization time series~\cite{Zhao2016}, that in case of phase-transition, there would be a remarkable change in scaling behavior measured in terms of DFA, Hurst exponent, MF-DFA parameter and complex network parameter.
Hence, in this work we have done the scaling analysis of the pseudorapidity data extracted from Pb-Pb VSD masterclass data sample at $2.76 TeV$ per nucleon pair from ALICE Collaboration~\cite{alice} using a completely different method of Visibility Graph from the perspective of complex network and also multifractal methodologies, to probe for phase transition and capture the signature of QGP.

The rest of the paper is organized as follows. The methods of analysis are described in Section~\ref{ana} - the Multifractal-Detrended Fluctuation Analysis(MF-DFA) method and the significance of parameters Hurst exponent and width of multifractal spectrum is described in Section~\ref{mfdfa},  Visibility Graph algorithm and the significance of its scale-freeness property, are presented in Section~\ref{vgalgo}. The data description is there in Section~\ref{data}. The details of our analysis and the inferences from the test results are given in Section~\ref{method}. The physical significance of the observable parameters and their significance with respect to the signature of QGP in multiparticle production process, is elaborated and the paper is concluded in Section~\ref{con}.

\section{Method of analysis}
\label{ana}
We would be describing Multifractal-Detrended Fluctuation Analysis(MF-DFA) method~\cite{kantel2002} to extract the Hurst exponent and the width of the multifractal spectrum and the Visibility graph technique~\cite{laca2008,laca2009} briefly in this section. We would be using these parameters for analyzing the fluctuation of pseudorapidity space($\eta$-space) of the event datasets extracted from the experimental data.
\subsection{MF-DFA method}
\label{mfdfa}
\begin{enumerate}
\item Let us denote the input data series as $x(i)$ for $i = 1,2,\ldots,N$, with $N$ number of points. The mean values of this series is calculated as $\bar{x} = \frac{1}{N}\sum_{i=1}^{N} x(i)$. Then accumulated deviation series for $x(i)$ is calculated as per the below equation.
\begin{eqnarray}
X(i) \equiv \sum_{k=1}^{i} [x(k)-\bar{x}], i = 1,2,\ldots,N  \nonumber 
\end{eqnarray}
This subtraction of the mean($\bar{x}$) from the data series, is a standard way of removing noisy data from the input data series. The effect of this subtraction would be eliminated by the detrending in the fourth step.

\item $X(i)$ is divided into $N_s$ non-overlapping segments, where $N_s \equiv int(N/s)$, $s$ is the length of the segment. In our experiment $s$ varies from $16$ as minimum to $1024$ as maximum value in log-scale.

\item For each $s$, we denote a particular segment by $v$($v = 1,2,\ldots,N_s$). For each segment least-square fit is performed to obtain the local trend of the particular segment~\cite{Peng1994}.
Here $x_v(i)$ denotes the least square fitted polynomials for the segment $v$ in $X(i)$. $x_v(i)$ is calculated as per the equations $x_v(i) = \sum_{k=0}^{m} {C_{k}}{(i)^{m-k}}$, where $C_{k}$ is the $k$th coefficients of the fit polynomials with degree $m$. For fitting linear, quadratic, cubic or higher $m$-order polynomials may be used~\cite{kantel2001,kantel2002}. For this experiment $m$ is taken as $1$ for linear fitting.

\item To detrend the data series, we have to subtract the polynomial fit from the data series. There is presence of slow varying trends in natural data series. Hence to quantify the scale invariant structure of the variation around the trends, detrending is required.
Here for each $s$ and segment $v \in 1,2,\ldots,N_s$, detrending is done by subtracting the least-square fit $x_v(i)$ from the part of the data series $X(i)$, for the segment $v$ to determine the variance, denoted by $F^2(s,v)$ calculated as per the following equation.
\begin{eqnarray}
F^2(s,v) \equiv \frac{1}{s}\sum_{i=1}^{s} \{X[(v-1)s+i]-x_v(i)\}^2, \nonumber
\end{eqnarray}
where $s \in 16,32,\ldots,1024$ and $v \in 1,2,\ldots,N_s$.

\item Then the $q$th-order fluctuation function, denoted by $F_q(s)$, is calculated by averaging $F^2(s,v)$ over all the segments($v$) generated for each of the $s \in 16,32,\ldots,1024$ and for a particular $q$, as per the equation below.
\begin{eqnarray}
F_q(s) \equiv \left\{\frac{1}{N_s}\sum_{v=1}^{N_s} [F^2(s,v)]^{\frac{q}{2}}\right\}^{\frac{1}{q}}, \nonumber
\end{eqnarray}
for $q \neq 0$ because in that case $\frac{1}{q}$ would blow up. In our experiment $q$ varies from $(-5)$ to $(+5)$. For $q = 2$, calculation of $F_q(s)$ boils down to standard method of Detrended Fluctuation Analysis(DFA)~\cite{cpeng1994}.

\item The above process is repeated for different values of $s \in 16,32,\ldots,1024$ and it can be seen that for a specific $q$, $F_q(s)$ increases with increasing $s$. If the series is long range power correlated, the $F_q(s)$ versus $s$ for a particular $q$, will show power-law behavior as below.
\begin{eqnarray}
F_q(s) \propto s^{h(q)} \nonumber
\end{eqnarray}
If this kind of scaling exists, $\log_{2} [F_q(s)]$ would depend linearly on $\log_{2} s$, where $h(q)$ is the slope which depends on $q$. $h(2)$ is similar to the well-known \textbf{Hurst exponent}~\cite{kantel2001}. So, in general, $h(q)$ is the generalized Hurst exponent.

\item For monofractal series, the scaling behavior of the variance $F^2(s,v)$ is exactly same for all segments. So, the averaging process would yield identical scaling behavior for different values of $q$ and so $h(q)$ becomes independent of $q$. 

But, if small and large fluctuations have different scaling behavior, then $h(q)$ becomes largely dependent on $q$. $h(q)$ describes scaling pattern of the segments with large fluctuations for positive values of $q$ and similarly, $h(q)$ describes scaling pattern of the segments with small fluctuations for negative values of $q$. So, the generalized Hurst exponent $h(q)$ for a multifractal series is related to the classical multifractal scaling exponent $\tau(q)$ as per the equation below.
\begin{eqnarray}
\tau(q) = qh(q)-1 \nonumber
\end{eqnarray}

\item Multifractal series have multiple Hurst exponents, and so $\tau(q)$ depends nonlinearly on $q$~\cite{Ashkenazy2003}. The singularity spectrum-$f(\alpha)$ is related to $h(q)$ as per the below equation.
\begin{eqnarray}
\alpha = h(q)+qh'(q), f(\alpha) = q[\alpha-h(q)]+1 \nonumber
\end{eqnarray}
Here $\alpha$ is singularity strength and $f(\alpha)$ describes the dimension of the subset series indicated by $\alpha$. The resultant multifractal spectrum $f(\alpha)$ is an arc where the difference between the maximum and minimum value of $\alpha$, is the \textbf{width of the multifractal spectrum} which is the amount of the multifractality of the input data series.

\end{enumerate}

\subsection{Visibility Graph Algorithm}
\label{vgalgo}

\begin{figure}[h]
\centerline{
\includegraphics[width=3in]{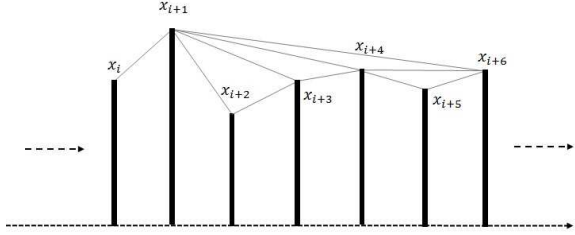}
}
\caption{Visibility Graph for time series X}
\label{visi}
\end{figure}
Visibility Graph algorithm plots time series $X$ to its Visibility Graph. Lets assume the $i^th$ point of time series is $X_{i}$. In this graph two nodes or vertices($X_{m}$ and  $X_{n}$) are supposed to be connected by a two-way edge if and only if the equation~\ref{ve} is valid.

\begin{equation}
X_{m+j} < X_{n} + (\frac{n - (m+j)}{n-m})\cdot(X_m - X_n) \nonumber
\label{ve}
\end{equation}
\begin{math}
\mbox{where }
\forall j \in Z^{+} \mbox{ and } j < (n-m)\\
\\
\end{math} 
It is shown in the Fig.~\ref{visi}, that the nodes $X_{m}$ and  $X_{n}$, where $m=i$ and $n=i+6$, are visible to each other only if the Eq. \ref{ve} is valid. 
It is evident that two sequential points of the time series can always see each other and thereby sequential nodes are always connected.

\subsubsection{Power of Scale-freeness of VG \mbox{-} PSVG}
\label{psvg}
The degree of a node or vertex in a graph \mbox{-} here Visibility Graph, is the number of connections or edges the node has with the rest of the nodes in the graph. The degree distribution $P(k)$ of a network is therefore defined as the fraction of nodes with degree $k$, with respect to the total number of nodes present in the network. So, if there are $n_k$ number of nodes in the network, having degree $k$ and total number of nodes in total in a network is $n$, then we define $P(k) = n_k/n$ for all possible values of $k$. 

As per Lacasa et al.\cite{laca2008,laca2009} and Ahmadlou et al.\cite{ahmad2012}, the degree of scale-freeness of a Visibility Graph corresponds to the amount of fractality and complexity of the time series.
According to the scale-freeness property of Visibility Graph, the degree distribution of its nodes should follow power-law, i,e, $P(k) \sim k^{-\lambda_p}$, where $\lambda_p$ is a constant and it is called the \textbf{Power of the Scale-freeness in Visibility Graph-PSVG}. 
Hence $\lambda_p$ or the PSVG corresponds to the amount of self-similarity, fractality and a measure of complexity of the time series. As the fractal dimension measures the amount of self-similarity of a time series, $\lambda_p$ indicates the FD \mbox{-} Fractal Dimension of the signal\cite{laca2008,laca2009,ahmad2012}. 
It is also observed that there is an inverse linear relationship between PSVG-$\lambda_p$ and Hurst exponent of the associated time series\cite{laca2009}.

\section{Experimental details}
\subsection{Data description}
\label{data}
In this experiment, we have extracted $10$ root dataset files, namely AliVSD\_Masterclass\_$1,2,\ldots,10$, each containing $34$ event datasets summing upto a total of $340$ event datasets from Pb-Pb VSD masterclass data sample at $2.76 TeV$ per nucleon pair from ALICE Collaboration~\cite{alice}. The derived dataset has been downloaded from :\href{http://opendata.cern.ch/record/1120}{\textit{CERN Open Data Portal}} which contains the above mentioned $10$ root dataset files. 
Derived datasets as defined in the glossary of CERN are the datasets which contain data that have been derived from the primary datasets. The data may be reduced in the sense that either only part of the information is kept or only part of the events are selected. In our case the $14$ primary datasets are also available at CERN Open Data Portal in this \href{http://opendata.cern.ch/search?page=1&size=20&type=Dataset&experiment=ALICE&subtype=Collision}{\textit{link}}.

\begin{figure*}
\centerline{\includegraphics[width=5in]{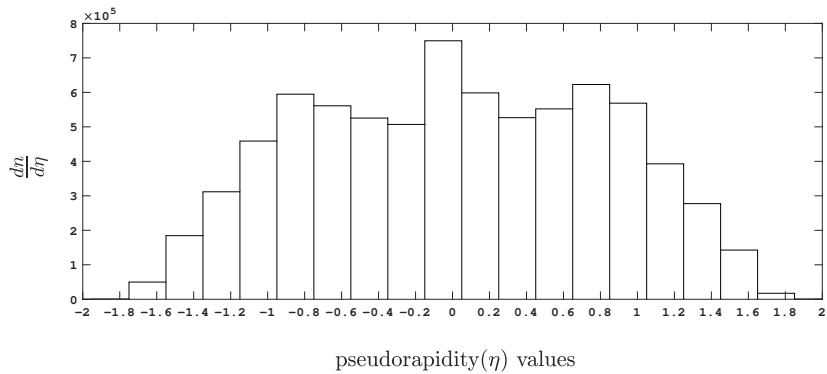}}
\vspace*{8pt}
\caption{The pseudorapidity distribution for Pb-Pb VSD masterclass data sample at $2.76 TeV$ per nucleon pair from ALICE Collaboration.\protect\label{eta_prob}}
\end{figure*}

Then from each of the $340$ event datasets, the pseudorapidity(denoted by $\eta$) space is extracted. Hence in effect we extracted $34$, $\eta$-space from each of the root dataset. Finally, total of $34 \times 10 = 340$ number of $\eta$-spaces are obtained. The distribution of the combined dataset of $\eta$-values of the $340$ datasets is shown in the Figure~\ref{eta_prob}

Then each of these $340$, $\eta$-space is created by putting $\eta$ values in a sequence. Along the $X$-axis the sequence number is plotted and each $\eta$-value corresponding to the sequence is plotted along the $Y$ axis. This way we have data series for each of the $340$, $\eta$-dataset. Also, one hadron-nucleus($\pi^{-}$-AgBr(350 GeV)) interaction dataset and one nucleus-nucleus($^{32}$S-AgBr(200 A GeV)) dataset extracted from an interaction with comparatively higher energy than the hadron-nucleus interaction. For $\pi^{-}$-AgBr(350 GeV) interaction the pseudorapidity($\eta$) coverage is $0.40<\eta<6.50$ and for$^{32}$S-AgBr(200 A GeV) interaction this coverage is $0.90<\eta<6.50$(lab system), the details of which are given in our earlier works~\cite{Bhaduri20164,Bhaduri20165,Bhaduri20166,Bhaduri20167,Bhaduri20171,Bhaduri20172,Bhaduri20183,Bhaduri20181,Bhaduri20182}. Each of these data series is analyzed using the methods described in Section~\ref{ana}. The method of analysis is described in detail in the Section~\ref{method}.

\subsection{Method of analysis and results}
\label{method}
\subsubsection{Calculation of Parameters}
\label{parm}
For each of the below datasets are analyzed
\begin{enumerate}
\item the $340$, $\eta$ datasets extracted from $340$ Pb-Pb VSD masterclass datasets at $2.76 TeV$ per nucleon pair from ALICE Collaboration~\cite{alice} and mapped into data series.
\item one hadron-nucleus($\pi^{-}$-AgBr(350 GeV)) interaction dataset and one nucleus-nucleus($^{32}$S-AgBr(200 A GeV)) dataset are also extracted. These are our own data and the details of these data are elaborated in Section~\ref{data}.
For both these datasets the full phase-space of the $\eta$ values extracted and mapped into data series. 
\end{enumerate}

For each of these $342=340+2$ data series below parameters are calculated.

\begin{itemize}
\item Hurst exponent
\item The width of the multifractal spectrum
\item PSVG-$\lambda_p$
\end{itemize}

The detailed steps of the calculating the parameters are described below.
For each of $342$, $\eta$-datasets, the $\eta$-values are plotted along the $Y$-axis in a sequence along the $X$-axis with equal interval. These way the data series from $\eta$-datasets are obtained.

\begin{itemize}
\item \textbf{\textit{Calculation of the Hurst exponent and the width of the multifractal spectrum:}}
Each one of these sequences was then subjected to the multi-fractal analyses described in sections Section~\ref{mfdfa} yielding $342$ values of the Hurst exponent, multi-spectral spectrum width.
Figure~\ref{hurst}, shows the linear trend of $\log_{2} s$ versus $\log_{2} [F_2(s)]$ calculated for one sample ALICE event dataset from one of the $340$ Pb-Pb VSD masterclass datasets at $2.76 TeV$ per nucleon pair from ALICE Collaboration. The value of the Hurst exponent is calculated from the slope of of $\log_{2} s$ versus $\log_{2} [F_2(s)]$ which $1.16\pm 0.01$ and the corresponding $\chi^2_{exp}$/DOF $=0.10$ value and the value of $R^2=0.96$.

\begin{figure*}
\centerline{\includegraphics[width=6in]{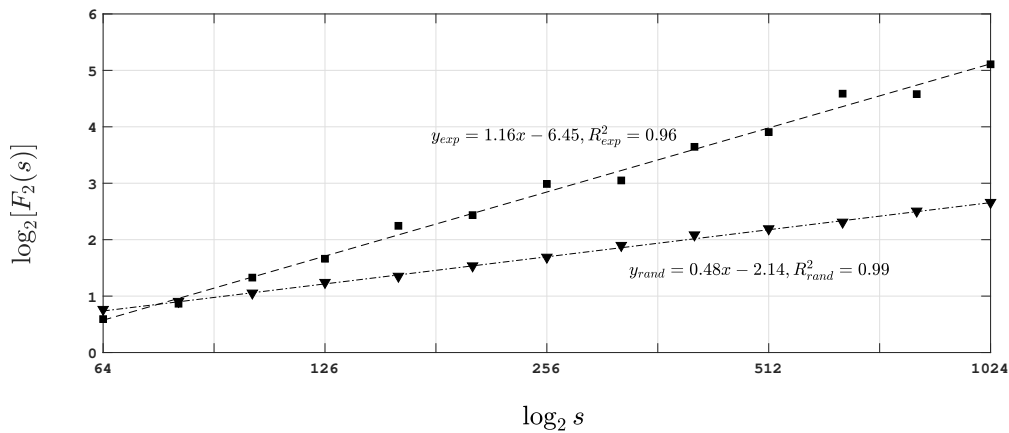}}
\vspace*{8pt}
\caption{$\log_{2} s$ vs $\log_{2} [F_2(s)]$ plot for the sample $\eta$-space of the same ALICE event dataset and its randomised version.\protect\label{hurst}}
\end{figure*}

Then, for each of the $342$, $\eta$-data series, randomized versions are generated and then their Hurst exponents and widths of the multifractal spectrum are calculated as per the process in Section ~\ref{mfdfa}. Then the values of the parameters are compared to those calculated for the experimental data. If the original data has long range correlations, that would be removed by the process of randomization and the data would become uncorrelated. So it's expected that the width of the multifractal spectrum and Hurst exponent for the randomized version would be remarkably different than those calculated for the experimental version. In the Figure~\ref{hurst}, the Hurst exponent of the randomized version of the same ALICE data series is shown, where the value is calculated as $0.48\pm 0.01$.

This has been found to be true for all the experimental $\eta$-data series extracted from the $340$ ALICE event datasets and two data series extracted from $\pi^{-}$-AgBr(350 GeV) and $^{32}$S-AgBr(200 A GeV) interaction, when compared with their randomized versions.

This establishes that the dynamics of pion fluctuations in the experimental data has not been the outcome of randomness inherent in the fluctuation pattern but rather is due to the broad probability distribution and long range correlation present in the data.

\item \textbf{\textit{Calculation of PSVG-$\lambda_p$:}}
\begin{enumerate}
\item 
For each of the $340$ sequences(for Alice) Visibility Graph is constructed as per the algorithm elaborated in Section~\ref{vgalgo} and for each of the Visibility Graphs $340$ values of Power of Scale-freeness of Visibility Graph \mbox{-} PSVG(denoted by $\lambda_p$) is extracted as per the method described in Section~\ref{psvg}.
We can understand that each of the node in the Visibility Graph constructed for each of the $340$ Alice datasets, is actually the value of the $\eta$-value(along $Y$-axis) in sequence along the $X$-axis.

For each of the Visibility Graphs, the $k$ vs $P(k)$ dataset is calculated as per the method described in Section~\ref{psvg}. $k$ vs $P(k)$ plot for the same ALICE event dataset of $\eta$-values, is shown in Figure~\ref{power_shower}-(a). The power-law index has been obtained by power-law fitting for the $k$ vs $P(k)$ datasets as per the method by Clauset et al.~\cite{Clauset2009}. The power-law relationship can be confirmed from the corresponding and the values of $R^2$.

\begin{figure*}
\centerline{
\includegraphics[width=3.5in]{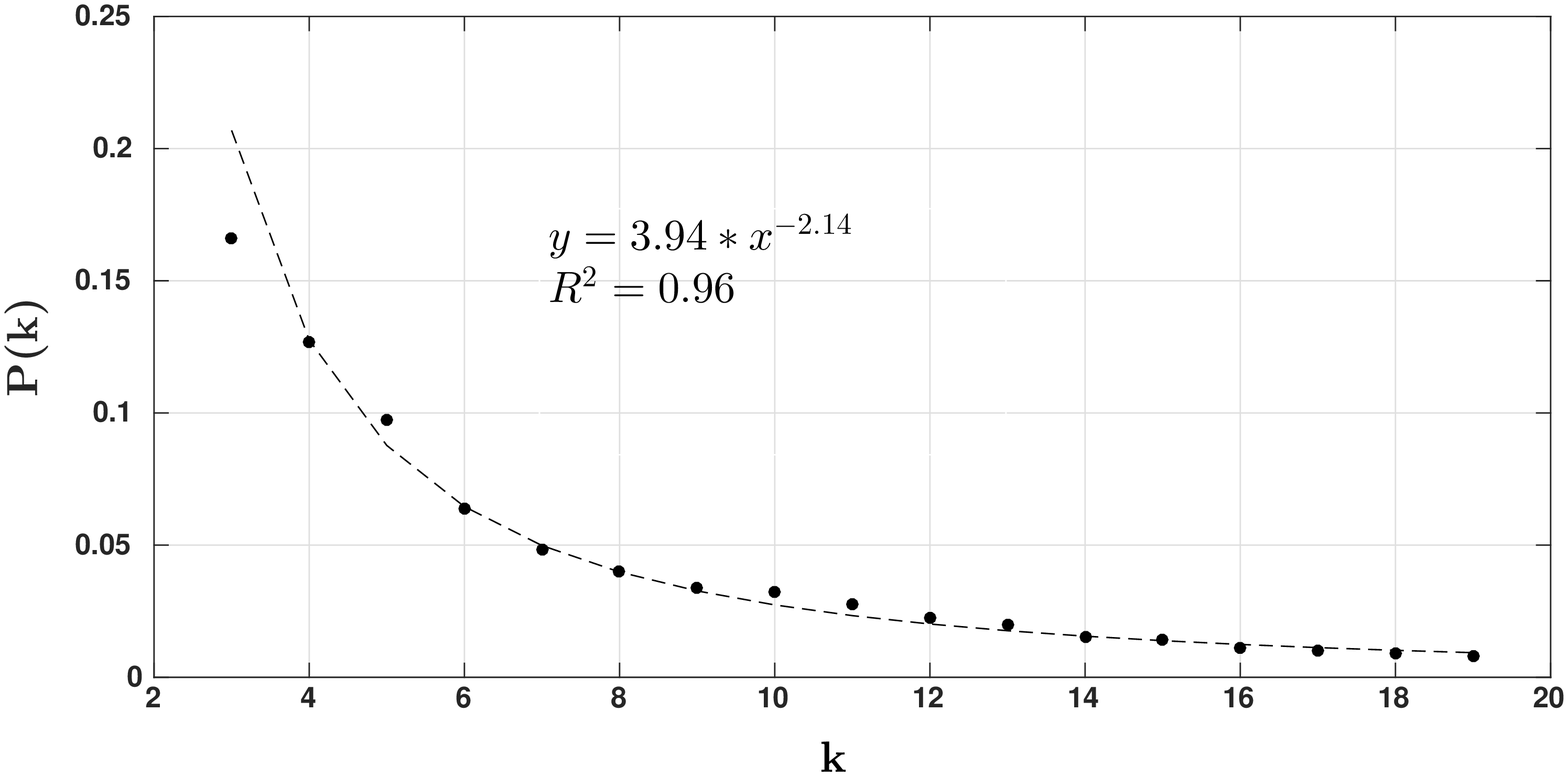}
\includegraphics[width=3.5in]{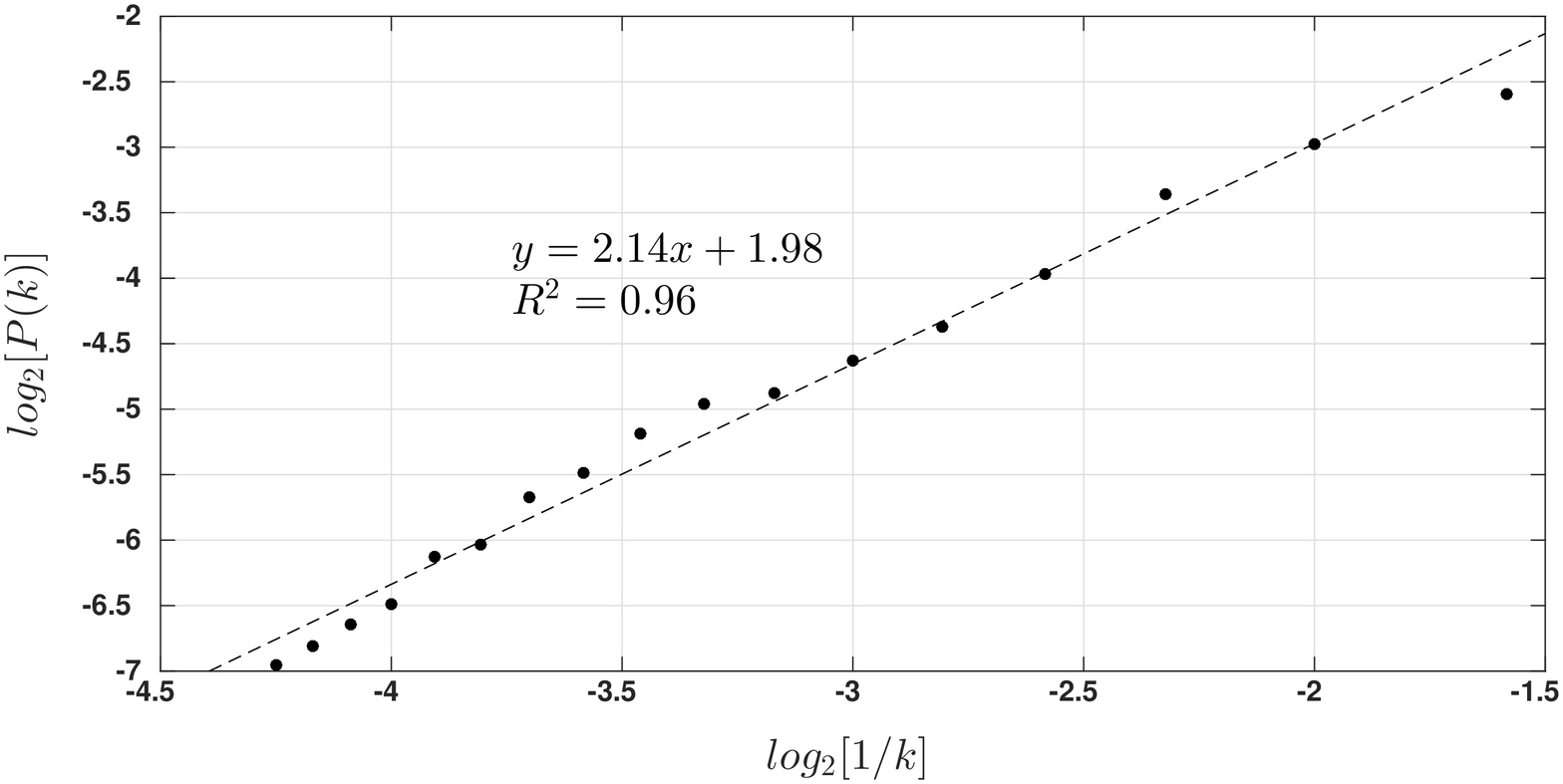}
}
\centerline{(a) \hspace*{6cm} (b)}
\caption{(a) $k$ vs $P(k)$ plot for $\eta$-space of a sample event dataset (b) $log_{2}[1/k]$ vs $log_{2}[P(k)]$ plot for the same $\eta$-space}
\label{power_shower}
\end{figure*}

The Power of Scale-freeness in Visibility Graph(PSVG), is calculated from the slope of $log_{2}[1/k]$ versus $log_{2}[P(k)]$ for all the Visibility Graphs constructed. It should be noted that PSVG corresponds to the amount of complexity and fractality of the data series and in turn indicates the fractal dimension of the data series~\cite{laca2008,laca2009,ahmad2012}.
Figure~\ref{power_shower}-(b) shows the plot of $log_{2}[1/k]$ versus $log_{2}[P(k)]$ for the same experimental dataset. $\lambda_p$ for this dataset is calculated as $2.14\pm 0.03$, the corresponding $\chi^2_{exp}$/DOF $=0.10$ value and the value of $R^2=0.96$. The same value for $\lambda_p$ has been obtained by power-law fitting done for the corresponding $k$ vs $P(k)$ dataset as shown in the Figure~\ref{power_shower}-(a).

Similar power-law trend is observed for the $k$ vs $P(k)$ datasets extracted from the rest of the $2$ datasets of $\pi^{-}$-AgBr(350 GeV) and $^{32}$S-AgBr(200 A GeV) interaction, apart from the ALICE event datasets. Also similar linear trend has been observed in the plot of $log_{2}[1/k]$ versus $log_{2}[P(k)]$, which is confirmed by the higher range of values for $R^2$.

\item Then we have repeated the Visibility Graph analysis and extracted PSVG values from the graphs constructed from the randomized version of each of the $342$ experimental $\eta$-datasets. 
Then, for each of the data series, we have compared PSVG values calculated for the - experimental and randomized version of data.

It has been found that values of $\lambda_{p}$ for the Visibility Graphs extracted from the experimental data series are remarkably different from those of the randomized versions for all of the $342$ data series of $\eta$-values. 
Hence it can be re-established that the dynamics of pion fluctuations in the experimental data for ALICE, $\pi^{-}$-AgBr(350 GeV) and $^{32}$S-AgBr(200 A GeV) interaction, is never the outcome of randomness inherent in the fluctuation pattern.

\end{enumerate}
\end{itemize}

This way, for each of the $342$ event datasets, a set of three values for each of the three parameters are obtained.

\subsubsection{Consolidating the Parameters Values for Comparison}
\label{conparm}
Most of the $340$ experimental event datasets of ALICE experiment are bigger is size than each of the $\pi^{-}$-AgBr(350 GeV) and $^{32}$S-AgBr(200 A GeV) interaction datasets. Few of them are similar in size. Hence the final parameter values for comparison are consolidated as per the following steps.
\begin{enumerate}
\item Then frequency histogram of $340$ values for each of parameters of PSVG-$\lambda_p$, Hurst exponent and the width of the multifractal spectrum is constructed for ALICE Pb-Pb interaction data. Then from each of the three histograms, the peak is extracted.
Each of the \textit{\textbf{peak}}-s of the 
\begin{itemize}
\item PSVG histogram is treated as the PSVG
\item Hurst exponent histogram is treated as the Hurst exponent
\item the multifractal spectrum-width histogram is treated as the width of the multifractal spectrum
\end{itemize}
respectively, for the ALICE Pb-Pb interaction data.

\item Two sets of three values of the parameters - PSVG-$\lambda_p$, Hurst exponent and the width of the multifractal spectrum are calculated for the two datasets of $\pi^{-}$-AgBr(350 GeV) and $^{32}$S-AgBr(200 A GeV) interaction are considered for comparison process.

\end{enumerate}

\subsubsection{Comparison of the Parameter Values}
\label{comparm}
Once the the values of the parameters - PSVG-$\lambda_p$, Hurst exponent and the width of the multifractal spectrum are consolidated for ALICE, $\pi^{-}$-AgBr(350 GeV) and $^{32}$S-AgBr(200 A GeV) interaction data as per the steps in Section~\ref{conparm}, they are compared among each other.
In the work~\cite{Bhaduri20167}, we have done scaling analysis of multiplicity fluctuation for hadron-nucleus($\pi^{-}$-AgBr(350 GeV)) and nucleus-nucleus($^{32}$S-AgBr(200 A GeV)) interaction with comparatively higher energy and have extracted the PSVG and other network parameters from the Visibility Graph. In this work, apart from PSVG, we have also extracted Hurst exponent and the width of the multifractal spectrum according to the method explained in Section~\ref{parm}, for the full phase-space of the $\eta$ values extracted from both the interaction datasets. 

In this work we have done comparison among the values of parameters - MF(Multifractal) Spectrum Width, PSVG and Hurst Exponent between ALICE Pb-Pb collision data at $2.76 TeV$ per nucleon pair, $\pi^{-}$-AgBr collision data at 350 GeV and $^{32}$S-AgBr collision data at 200 A GeV data, calculated as per the method explained in Section~\ref{parm} and the consolidated as per steps described in Section~\ref{conparm}. The comparison is shown in Table~\ref{comp_table}.



\begin{table*}
\caption{Comparison of the experimental values of MF Spectrum Width, PSVG and Hurst Exponent respectively, among ALICE Pb-Pb data at $2.76 TeV$ per nucleon pair, $\pi^{-}$-AgBr(350 GeV) and $^{32}$S-AgBr(200 A GeV) data, with their randomized version.}
\begin{tabular}{@{}|l|c|c|c|c|c|c|@{}} \hline
&\multicolumn{2}{|c|}{\textbf{ALICE Pb-Pb($2.76 TeV$)}}&\multicolumn{2}{|c|}{\textbf{$^{32}$S-AgBr($200 A GeV$)}}&\multicolumn{2}{|c|}{\textbf{$\pi^{-}$-AgBr($350 GeV$)}} \\
\cline{2-7}
&\textbf{Experimental}&\textbf{Random}&\textbf{Experimental}&\textbf{Random}&\textbf{Experimental}&\textbf{Random}\\ \hline
PSVG&$2.16$&$3.25$&$3.17$&$3.68$&$3.29$&$3.12$\\
Hurst exponent&$1.13$&$0.45$&$0.78$&$0.53$&$0.59$&$0.46$\\
MF Spectrum Width&$1.18$&$0.05$&$0.46$&$0.07$&$0.65$&$0.10$\\
\hline
\end{tabular}\label{comp_table} 
\end{table*}


The Table~\ref{comp_table} shows that the change of values for all the three parameters for either of $\pi^{-}$-AgBr(350 GeV) data or $^{32}$S-AgBr(200 A GeV) data to those consolidated for the the ALICE Pb-Pb data at $2.76 TeV$ per nucleon pair is \textit{remarkably} more than the change of the values of the parameters from $\pi^{-}$-AgBr(350 GeV) data to $^{32}$S-AgBr(200 A GeV) data.
One can infer that this remarkable change of all these $3$ parameters calculated by two different methods for analysis from $\pi^{-}$-AgBr(350 GeV) and $^{32}$S-AgBr(200 A GeV) interaction to ALICE Pb-Pb interaction data at $2.76 TeV$ per nucleon pair, definitely indicates the presence of higher long-range correlations in ALICE data. Since, remarkably different value of long-range correlation indicates phase transition~\cite{Zebende2004,Zhao2016}, this finding may be interpreted as a clear signature of QGP.

\section{Conclusion} 
\label{con}
The study of fluctuation in terms of an important observable - $\eta$ in the process of pionisation in ALICE Pb-Pb interaction data at $2.76 TeV$ per nucleon pair, in both multifractal and complex network scenario, presents that how can we obtain the signature of QGP~\cite{alice}, by analyzing the fluctuation pattern using robust, rigorous and also novel methodologies. The observations we got from the analysis are astonishing and listed below.
\begin{enumerate}
\item $340$ number of $\eta$-space values extracted from corresponding event dataset of the Pb-Pb VSD masterclass dataset at $2.76 TeV$ per nucleon pair from ALICE Collaboration~\cite{alice}, one $\eta$-dataset of $\pi^{-}$-AgBr(350 GeV) and one of $^{32}$S-AgBr(200 A GeV) interaction, are analyzed using a completely different method of Visibility Graph from complex network perspective and it has been shown that the fluctuation process of each $\eta$-space obeys scaling laws and hence confirms its fractality. The scaling behavior of a sample ALICE event dataset has been shown in Figure~\ref{power_shower}.

\item Also, multifractal analysis is done for each of the experimental dataset and Hurst exponent and the width of the multifractal spectrum are extracted for each of the $\eta$ dataset. The linear trend of $\log_{2} s$ versus $\log_{2} [F_q(s)]$ calculated for each $\eta$-dataset re-establishes scale-freeness and fractality inherent in the fluctiuation pattern. Figure~\ref{hurst} shows the linear trend of $\log_{2} s$ versus $\log_{2} [F_2(s)]$ for the same sample dataset.

\item Comparison of PSVG, Hurst exponent and the width of the multifractal spectrum extracted from experimental dataset with those of the randomized version of the experimental datasets reveals that pattern of pion fluctuations in the experimental data has never been the outcome of randomness inherent in the fluctuation pattern but the result of the broad probability distribution and long range correlation present in the data.

\item We have compared the values of all the three parameters calculated and consolidated for ALICE Pb-Pb interaction data at $2.76 TeV$ per nucleon pair with $\pi^{-}$-AgBr(350 GeV) and $^{32}$S-AgBr(200 A GeV) interaction data as per the method explained in Section~\ref{comparm}. The Table~\ref{comp_table} shows that there has been remarkable change in the values of the parameters from ALICE data to the values calculated for both the $\pi^{-}$-AgBr(350 GeV) and $^{32}$S-AgBr(200 A GeV) interaction data, which obviously indicates higher long-range correlations observed in ALICE data.

It should also be noted in the Table~\ref{comp_table} that the trend of the change of Hurst exponent among the interactions is just the \textit{reverse} of that of PSVG parameter. This is in conformance to the inverse linear relationship between PSVG and Hurst exponent of the data series\cite{laca2009} in question.

\item As remarkably different value of long-range correlation indicates phase transition~\cite{Zebende2004,Zhao2016}, similar change in the fluctuation pattern in terms of three rigorous parameters from ALICE Pb-Pb interaction data at $2.76 TeV$ per nucleon pair compared to $\pi^{-}$-AgBr(350 GeV) and $^{32}$S-AgBr(200 A GeV) interaction data can also be attributed to a phase transition and also the onset of QGP.
\end{enumerate}

This new approach promised to add another powerful tool based on non-statistical pion fluctuation to capture the signature of QGP apart from other state of the art methods.
\section*{Acknowledgements} 
\label{ack}
We thank the \textbf{Department of Higher Education, Govt. of West Bengal, India} for logistics support of computational analysis.

\section{References}

\end{document}